\documentstyle[12pt]{article}
\def\overlay#1#2{\ifmmode \setbox 0=\hbox {$#1$}\setbox 1=\hbox to\wd 0{\hss
$#2$\hss }\else \setbox 0=\hbox {#1}\setbox 1=\hbox to\wd 0{\hss #2\hss }\fi
#1\hskip -\wd 0\box 1}

\begin{document}

\font\fortssbx=cmssbx10 scaled \magstep2
\hbox to \hsize{
\hskip.5in \raise.1in\hbox{\fortssbx University of Wisconsin - Madison}
\hfill$\vtop{\hbox{\bf MAD/PH/777}
                \hbox{\bf DTP/93/58}
                \hbox{\bf RAL-93-054}
                \hbox{July 1993}}$ }

\vspace{.5in}

\begin{center}
{\large\bf Event shape criteria for single-lepton top signals}\\
[.4in]
V.~Barger$^a$, J.~Ohnemus$^b$ and R.J.N.~Phillips$^c$\\[.2in]
{\it $^{a}$Physics Dept., University of Wisconsin, Madison WI 53706, USA\\
$^{b}$Physics Dept., University of Durham, Durham, DH1 3LE, England\\
$^{c}$Rutherford Appleton Laboratory, Chilton, OX11 0QX, England}
\end{center}

\vspace{1in}

\begin{abstract}
\normalsize
Single-lepton plus jets signals from $t\bar t$ production at hadron
colliders generally give more spherically symmetrical events
than the principal backgrounds from $W$ production.  We show that
sphericity and aplanarity criteria, applied to lepton plus neutrino plus four
jet final states at the Fermilab Tevatron $p\bar p$ collider, help to
discriminate;
they can be used either to validate an eventual top signal or simply to reduce
background. An alternative circularity criterion in the transverse plane is
less successful.

\end{abstract}

\thispagestyle{empty}

\newpage

   The predicted top quark, an essential component of the Standard
Model (SM), is apparently too heavy to have been discovered yet at high
energy colliders; the present experimental limit is $m_t > 113$~GeV from the
CDF detector~\cite{cdf} and independently $m_t>103$~GeV from the D0
detector~\cite{D0}.  The consistency of radiative corrections to all
electroweak data however indicates that  $m_t = 150 {+19\atop-24}
{+15\atop-20}$~GeV~\cite{langacker}, and hence that
the top quark can probably be discovered at the Fermilab Tevatron
$p\bar p$ collider sooner or later~\cite{proto}. Here the SM predicts
mainly $t\bar t$ pair production via QCD with $t \to bW$ decays; the smallest
and cleanest signals are expected in two-lepton channels~\cite{bbop,han},
but larger single-lepton~\cite{bbp,berends} and all-jet~\cite{giele}
signals can also in principle be separated from backgrounds. Since the
top signals will initially (and perhaps for a considerable time) be
based on a small number of events, it will be important to suppress
the backgrounds as far as possible in ways that do not seriously reduce
the signals.  In this note we propose the use of event-shape criteria
to distinguish the top signal and reduce backgrounds in the relatively copious
single-lepton-plus-4\,jet channel.

   The underlying idea is that Tevatron $t\bar t$ production will be mainly
near threshold in the CM frame of the lowest-order QCD subprocesses
$q\bar q \to t\bar t$ or $gg \to t\bar t$, and hence will mostly lead to
rather spherical event shapes in this frame.  Phase-space factors
favor spherical configurations; the $t\bar t$ spin correlations~\cite{spincor}
and decay matrix elements introduce substructure but do not drastically
change this spherical tendency, which can help to distinguish the signal
from backgrounds.   This idea is very familiar in the context
of heavy quark searches at $e^+ e^-$ colliders and has recently been
applied to 6-jet top signals at the Tevatron~\cite{giele}; we propose
here to apply it to lepton-neutrino-4\,jet signals from $t\bar t \to
b\bar bWW \to b\bar b q\bar q' \ell\nu$ decays (where the neutrino is
partially measured via transverse energy balance).  Backgrounds in this
channel from $b$- and $c$-quark semileptonic decays can be removed by
lepton $p_T$ and isolation cuts~\cite{bbp}. The major background then comes
from $W$ production plus 4 QCD jets, which remains comparable with the signal
after all the usual cuts on transverse momenta $p_T$, pseudo-rapidities
$\eta$, and missing transverse energy $\overlay /E_T$~\cite{bbp,berends}; we
neglect contributions from $WW$ or $WZ$
plus 2 QCD jets~\cite{ww} that are an order
of magnitude smaller.
This background can be further suppressed by requiring a $b$-jet-tag,
costing the signal a tagging-efficiency factor of order 0.3 at
present~\cite{proto}.
Alternatively, we may expect the inherent collinear singularities to
cause spatial correlations of QCD jets with each other and with the
beam axis, giving less spherical configurations and hence a discrimination
between background and signal via an appropriate choice of event-shape
variable as follows.

   First we restrict ourselves to momenta in the transverse plane,  in order to
use the most direct neutrino information contained in its transverse momentum
defined by ${\bf p}_T(\nu) = {\overlay/{\bf E}}_T$. We consider the usual
``circularity" event shape variable $C$ defined by
\begin{equation}
C = 2 \min { \sum_i ({\bf p}_T^i \cdot \hat{{\bf n}} )^2 \over
             \sum_i (p_T^i)^2 } \,,
\end{equation}
summed over the lepton, neutrino, and 4~jet transverse momenta and minimized
with respect to the choice of the unit vector $\hat{\bf n}$.  This quadratic
variable is unstable against the splitting or combining of
jets; although $C$ is well defined for any given jet algorithm, it would be
theoretically preferable to use a linearized variable such as
\begin{equation}
C' = {\pi \over2} \min { \sum_i \bigl|{\bf p}_T^i \cdot \hat{{\bf n}} \bigr|
                  \over  \sum_i \bigl|p_T^i\bigr| } \,.
\end{equation}
Both $C$ and $C'$ are normalized to 1 for an ideal circular event and vanish
for a linear configuration.

   Next we consider event shapes in three dimensions.  The longitudinal
neutrino momentum $p_L(\nu)$ is not measured; it can be inferred
approximately by requiring that the lepton-neutrino invariant mass
$m(\ell\nu) = M_W$, but this leaves a two-fold ambiguity.
We choose the solution
that allows the most consistent $t\bar t$ reconstruction (details below) and
form the conventional normalized $3\times3$ momentum tensor
\begin{equation}
M_{ab} = { \sum_i p_a^i p_b^i \over \sum_i (p^i)^2 }
\end{equation}
from the lepton plus neutrino plus 4--jet momenta $p_a^i\ (i=1,\dots6;
a=1,2,3)$, in the CM frame of these momenta.
$M_{ab}$ has three eigenvalues $Q_i$ with  $0 \leq Q_1 \leq Q_2 \leq Q_3$ and
$Q_1 + Q_2 + Q_3 = 1$; the sphericity $S$ and aplanarity $A$ are then defined
by $S = {3\over2}(Q_1 + Q_2)$ and $A = {3\over2}Q_1$. (It would be
theoretically  preferable to use linearized variables such as thrust
and acoplanarity, but for ease of computation we use $S$ and $A$ here).
Thus $S=2A=1$ for ideal
spherical events, $S={3\over4}$ and $A=0$ for plane circular events, and
$S=A=0$ for linear events.   We therefore expect the Tevatron top signal to
give typically larger $C,\ C',\ S,$ and $A$ than the $W+{}$jets background, and
have made quantitative parton-level calculations at $\sqrt s = 1.8$~TeV to
confirm this.

For illustration we compute the $p\bar p \to t\bar tX$ production and
decay distributions in lowest order with full spin correlation
effects~\cite{spincor},
using the MRS set $S_0$ parton distributions of
Ref.~\cite{partons} evaluated at scale
$Q=m_t$, neglecting possible
additional QCD jets and taking $\ell + \nu + 4\,$jet final states.
The total $t\bar t$ production rate is normalized
conservatively to the central value of Ref.~\cite{ellis},
although values about 20\% higher have recently been proposed~\cite{laenen}.
We calculate the $W + 4\,$jet background from the full tree-level matrix
elements following Ref.~\cite{berends}, evaluating the parton distributions at
scale $Q=\left<p_T\right>$ in accord with Tevatron
results~\cite{abe1}.  We sum lepton flavors and
charges $\ell = e^{\pm},\mu^{\pm}$.  We simulate calorimeter errors
by gaussian uncertainties on jet  and lepton
energies, following the CDF values tabulated in Ref.~\cite{abe2}, and calculate
the apparent neutrino $p_T$ from overall $E_T$ imbalance.   We impose the
following acceptance cuts on transverse momenta, pseudorapidities, and jet
multiplicity $N_j$,
selecting regions where the top signal should be strong,
\begin{equation}
\begin{array}{rl@{\qquad}rl@{\qquad}rl}
p_T(j)    &> 20\rm\ GeV \,, & |\eta(j)|    &< 2 \,,& N_j &= 4\,,  \\
p_T(\ell) &> 20\rm\ GeV \,, & |\eta(\ell)| &< 2 \,,& \overlay/ E_T &> 20 \rm\
GeV\,.
\end{array} \label{accept}
\end{equation}
We also set minimum jet-jet and lepton-jet separations
\begin{equation}
   \Delta R(j,j) > 0.7  \,, \qquad \Delta R(\ell, j) > 0.7 \,, \label{sep}
\end{equation}
where $(\Delta R)^2 = (\Delta \eta)^2 + (\Delta \phi)^2$, to mimic some
effects of jet-finding algorithms and lepton isolation cuts~\cite{abe1}.
We note that these $\Delta R$ cuts already introduce some event-shape
discrimination.

We first compare the shapes of the $C,\ C',\ S,$ and $A$ distributions for
$t\bar t$ signal and $W +$~jets background, for the case $m_t=150$~GeV.  The
means $\mu$ and standard deviations $\sigma$  for the signal (background)
are
\begin{equation}
\begin{array}{rl@{\qquad}rl}
\mu(C) &= 0.51\ (0.48)\,, &  \sigma(C) &= 0.21\ (0.22) \,,\\
\mu(C') &= 0.68\ (0.66)\,, &  \sigma(C') &= 0.14\ (0.15) \,,\\
\mu(S) &= 0.50\ (0.41)\,,  & \sigma(S) &= 0.18\ (0.18) \,,\\
\mu(A) &= 0.135\ (0.090)\,, & \sigma(A) &= 0.079\ (0.066),
\end{array}
\end{equation}
from which we see that the circularity variables offer little
discrimination in this case, whereas the $S$ and $A$ variables show
appreciable displacements between signal and background. The $S$ and $A$
distributions are shown in Fig.~\ref{S,A dist}.   We see that the signal has
significantly different distributions
from the background, expecially for $A$; this could
help to validate an eventual top signal (extracted by $b$-tagging say).
Alternatively, cutting out events
below a minimum value $S_{\min}$ or $A_{\min}$ would reduce the background more
than the signal; integrated cross sections for $S > S_{\min}$ and $A >
A_{\min}$ are shown in Fig.~\ref{sigma min}.
We shall illustrate below the effects of a choice $S_{\min} = 0.20,\
A_{\min} = 0.05$ (not specifically optimized), similar to cuts suggested for
all-jet signals~\cite{giele}; this reduces the signal
(background) by 15\% (36\%), giving a cleaner result and an increase in its
statistical significance
(signal)/$\sqrt{\rm(background)}$.

   To extract a top quark mass peak, we introduce a constrained
event parameterization as follows.  We first compute the two solutions for
$p_L(\nu)$ by requiring  $m(\ell\nu) = M_W$; if they are formally complex,
we set the imaginary parts to zero, getting the single
nearest-to-onshell solution.  We then select the pair of jets with
invariant mass closest to $M_W$, and identify them as the $W \to jj$
decay candidates.  The remaining two jets are candidates for $b$ and $\bar b$
jets, if this is a $t\bar t$ event. From the four possible ways of pairing the
latter two jets with our unique $W \to jj$ choice and the two-fold
$W \to \ell \nu$ solution, we select whichever pairing gives the
closest agreement between the resultant $b+W$ invariant masses.  The mean
of these two closest $m(bW)$ values is defined as the reconstructed top
mass $\tilde m_t$ (following the notation of Ref.~\cite{bbop}); taking the mean
gives a narrower signal peak than either the semileptonic or the hadronic
$m(bW)$ value separately.  Our $\tilde m_t$ variable differs from those
proposed in Refs.~\cite{bbp,berends}.
Figure~\ref{tilde m} presents the calculated $\tilde m_t$ distribution
for signal and background, for $m_t = 130$, 150, and 170~GeV, first with the
standard acceptance cuts of Eqs.~(\ref{accept})--(\ref{sep}) (upper curves),
and also with the additional event-shape cuts $S > 0.20,\ A > 0.05$ (lower
curves). Hadronization effects have not been included, except
in their contributions to calorimeter resolution; we
expect they will somewhat further smear the  $\tilde m_t$ peak.

Our results show that
\begin{enumerate}
\item[(a)] $S$ and $A$ distributions of top signals differ significantly from
   background in the lepton + neutrino + 4\,jets channel;
\item[(b)] these differences could help to validate an eventual top signal
   or alternatively reduce background and improve significance;
\item[(c)] the signal would peak in our reconstructed top mass variable
   $\tilde m_t$.
\end{enumerate}
Event shape criteria have previously proved promising in the 6\,jet
channel~\cite{giele}; they may also have value in lepton-plus-3 jet
and dilepton channels.

\section*{Acknowledgments}

We thank Walter Giele for providing a copy of the VECBOS program for $W+4\,$jet
production and thank Tao Han for discussions. JO would like to thank the
Rutherford Appleton Laboratory Theory Group and the DESY Theory Group for
hospitality during the course of this work. JO is supported by the UK Science
and Engineering Research Council. The work of VB is supported in part by the
U.S.~Department of Energy under Contract No.~DE-AC02-76ER00881, in part by the
Texas National Laboratory Research Commission under Grant No.~RGFY93-221, and
in part by the University of Wisconsin Research Committee with funds granted by
the Wisconsin Alumni Research Foundation.

\newpage

\section*{Figure Captions}

\begin{enumerate}

\item\label{S,A dist}
Sphericity $S$ and aplanarity $A$ distributions for $t\bar t$ signal
($m_t=150$~GeV) and $W+{}$jets background, for $\ell + \nu + 4$\,jet events at
the Tevatron.

\item\label{sigma min}
Integrated signal and background cross sections for (a)~$S>S_{\min}$
and (b)~$A>A_{\min}$, versus $S_{\min}$ and $A_{\min}$, respectively.

\item\label{tilde m}
Reconstructed top-quark mass $\tilde m_t$ distributions for Tevatron $t\bar t$
signals with $m_t=130$, 150, 170~GeV, and $W+4$\,jets background. Upper
curves are for $S>0,\ A>0$; lower curves are for $S>0.20,\ A>0.05$.

\end{enumerate}

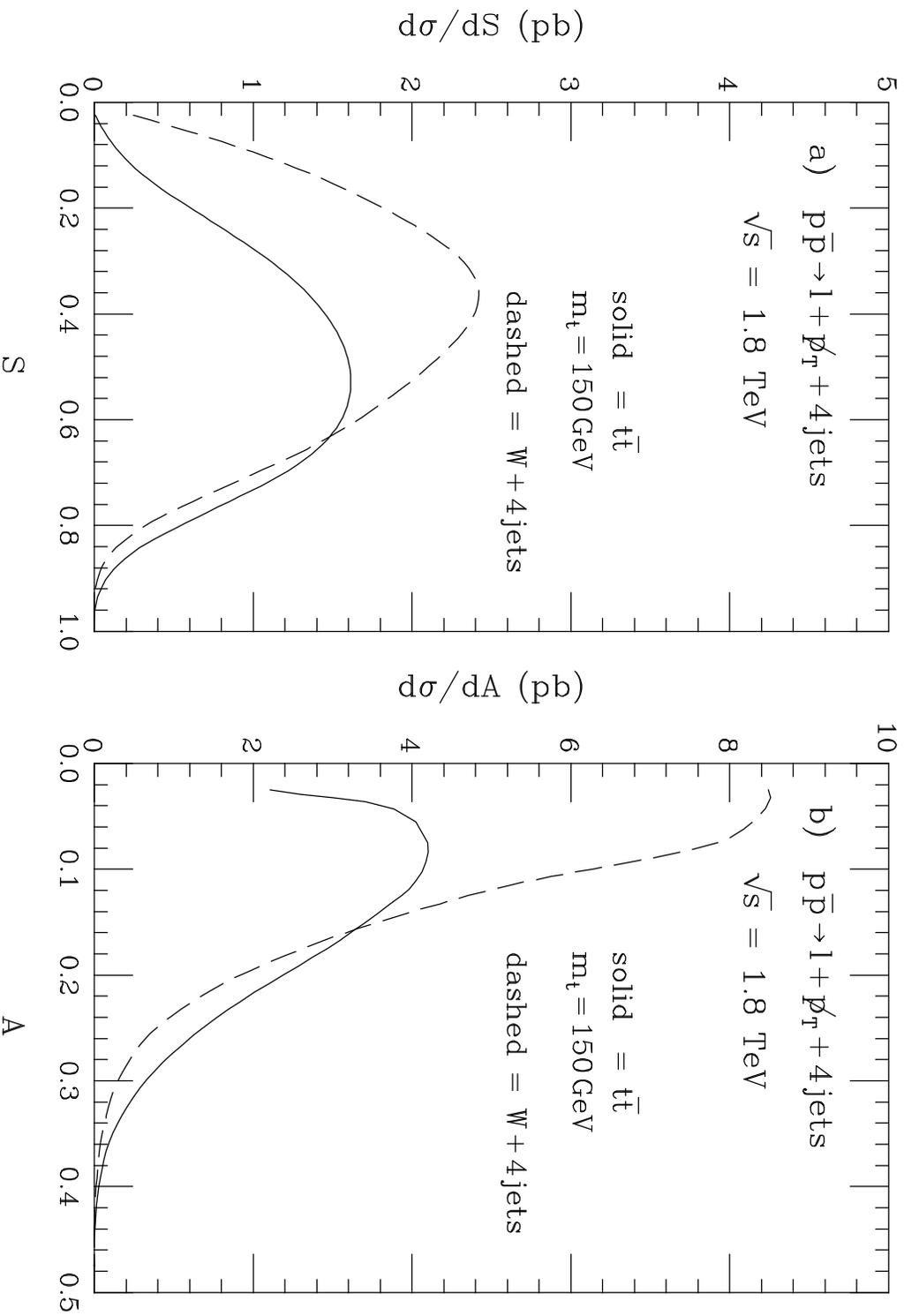
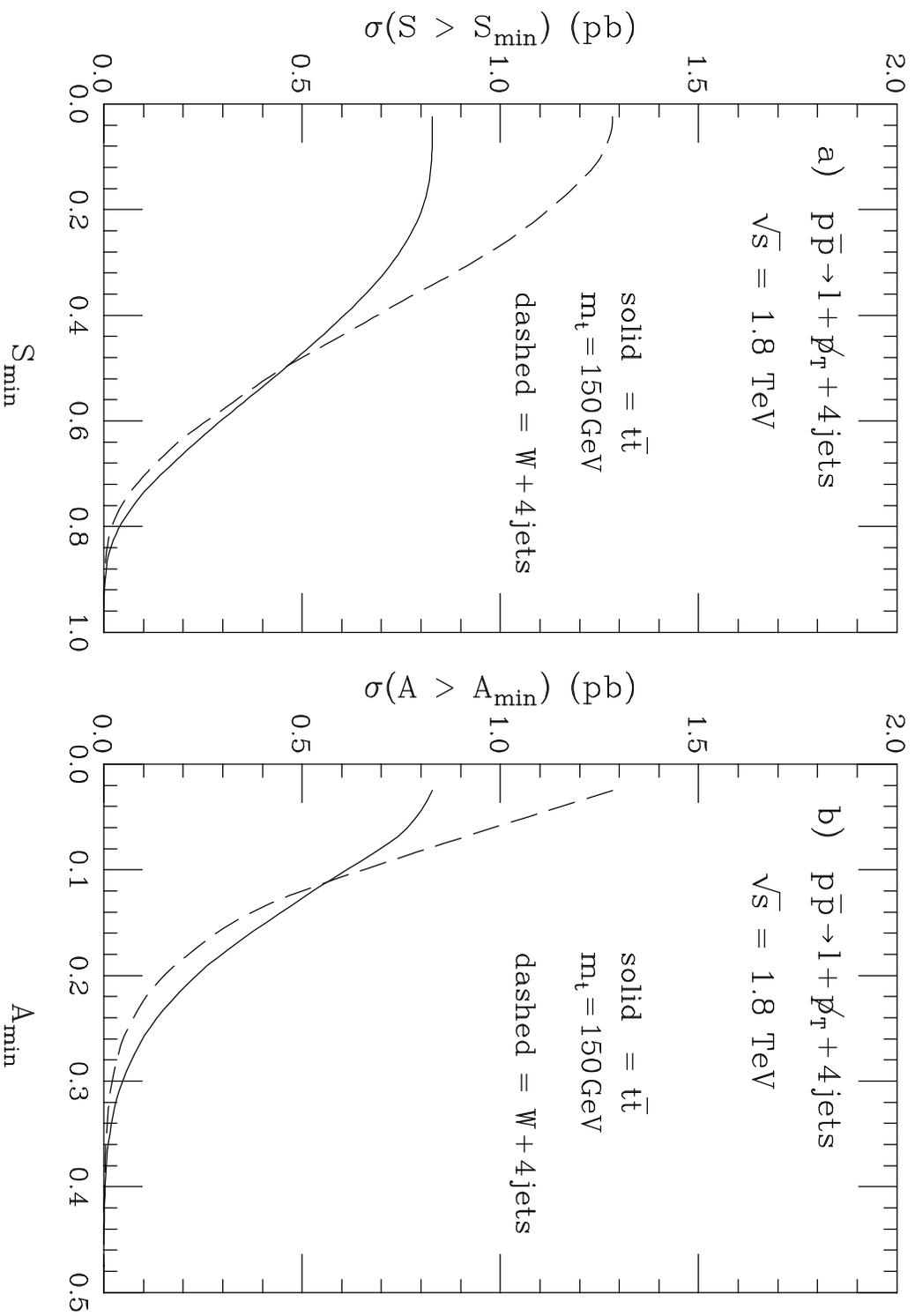
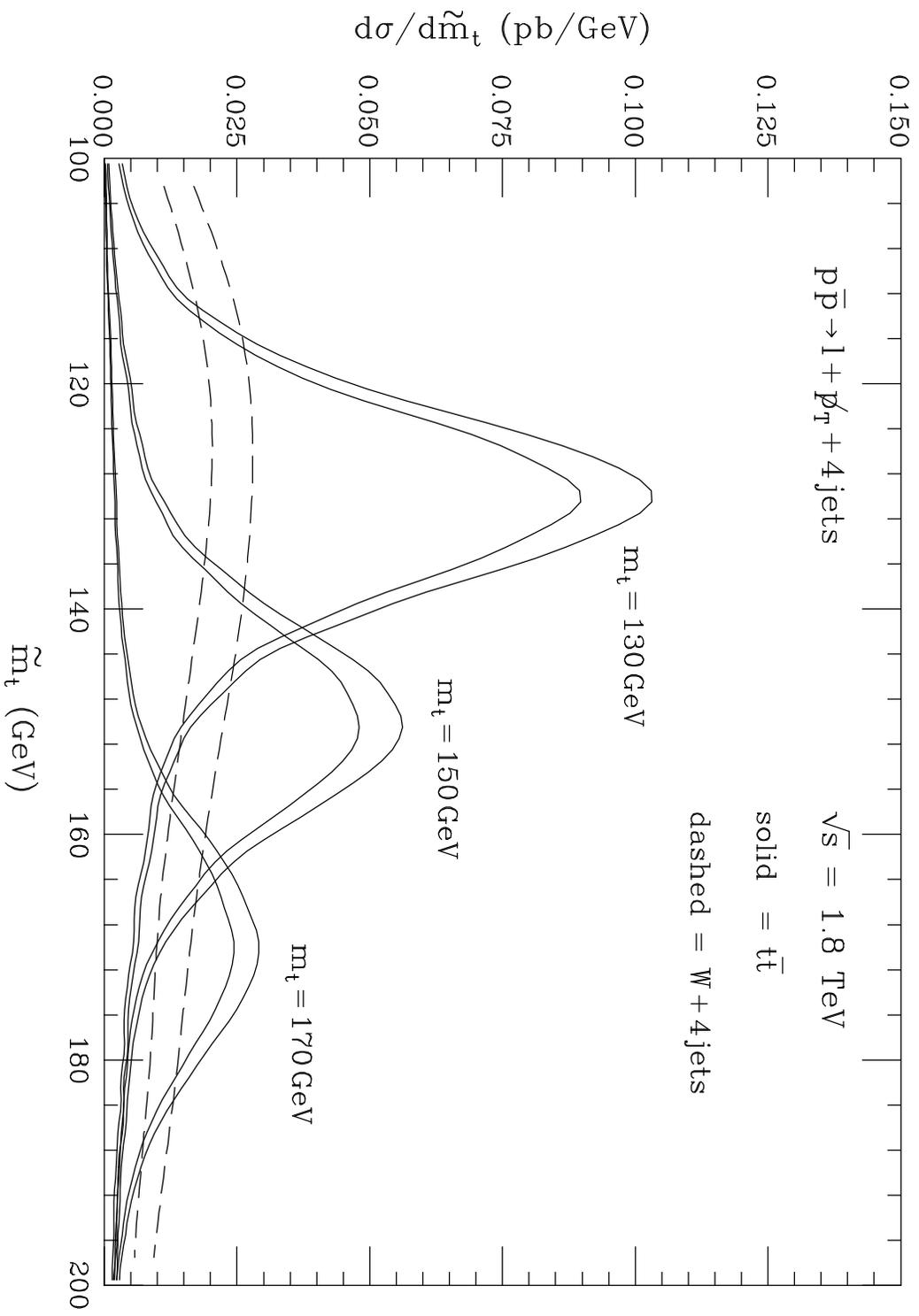
\end{document}